\documentclass[a4paper]{amsart}

\usepackage{amsmath}
\usepackage{fix-cm}
\usepackage{graphicx}
\usepackage{hyperref}
\usepackage{amsthm}
\usepackage{amssymb}
\usepackage{bm}
\usepackage{verbatim}
\usepackage{inputenc}
\usepackage{cite}
\usepackage[a4paper,left=4cm,width=13cm,right=3cm,top=3cm,bottom=3cm]{geometry}

\newtheorem{theorem}{Theorem}

\begin{document}

\title[Solutions of Inverse Problem for Vlasov-Maxwell Equilibria ]{Particular solutions of the Inverse Problem for 1D Vlasov-Maxwell equilibria using Hermite polynomials}

\author{O. Allanson, T. Neukirch, S. Troscheit and F. Wilson}
\address{School of Mathematics and Statistics, University of St Andrews, North Haugh, St Andrews, KY16 9SS, UK}
\email{oliver.allanson@st-andrews.ac.uk}

\begin{abstract}
We present the solution to an inverse problem arising in the context of finding a distribution function for 
a specific collisionless plasma equilibrium. The inverse problem involves the solution of two integral equations, each having the form of a Weierstrass transform. We prove that inverting the Weierstrass transform using Hermite polynomials leads to convergent infinite series. We also comment on the non-negativity of the distribution function, with more detail on this in Allanson \emph{et al., Journal of Plasma Physics}, vol. 82 (03), 2016. Whilst applied to a specific magnetic field, the inversion techniques used in this paper (as well as the derived convergence criteria and discussion of non-negativity) are of a general nature, and are applicable to other smooth pressure functions.
\end{abstract}
\date{\today}
\maketitle

\section{Introduction}
Plasma equilibria are of great importance for laboratory, space and astrophysical applications, see \cite{Krallbook-1973, Schindlerbook} for example. In this paper we present results arising from an inverse problem in the context of collisionless and non-relativistic plasma equilibria, namely that of calculating the 1-particle distribution function, given a prescribed magnetic field. We consider a system depending only on one Cartesian spatial coordinate $(x, y, z)$, namely the $z$-coordinate. In line with previous work
\cite{Channell-1976,Harrison-2009a,Neukirch-2009}, we will make several simplifying assumptions.
Firstly, we assume a two species plasma of ions and electrons with equal and opposite charge of magnitude $e$, such that the charge density vanishes. Then the electric field $\mathbf{E}$ vanishes, and Poisson's equation is automatically satisfied. A second assumption we make is that the magnetic field $\mathbf{B}$ only has two non-vanishing components, namely that $\mathbf{B}(z)=(B_x(z), B_y(z), 0)$. For an equilibrium, one then has to solve the coupled Vlasov-Maxwell equations
\begin{eqnarray}
v_{z} \frac{\partial f_s}{\partial z} + \frac{q_s}{m_s} (\mathbf{v} \times \mathbf{B} )\cdot \nabla_{\mathbf{v}}\ f_s = 0, \label{eq:vlasov} \\
\frac{d^2 A_x}{d z^2} = \mu_0 j_x = \mu_0 \sum\limits_s q_s \int v_{x}\, f_s\, d^3 v, \label{eq:ampere-x} \\
\frac{d^2 A_y}{d z^2} = \mu_0 j_y = \mu_0 \sum\limits_s q_s \int v_{y}\, f_s\, d^3 v \label{eq:ampere-y},
\end{eqnarray}
where $f_s(z,v_{x},v_{y},v_{z})$ $( \ge 0)$ is the 1-particle distribution function (DF) for particle species $s$ with charge $q_s$ and mass $m_s$, defined over a $4D$ phase-space of $z$ and $\mathbf{v}$. We have also introduced the vector potential $\mathbf{A}(z) = ( A_x(z), A_y(z), 0)$ with $\mathbf{B} = \nabla \times \mathbf{A}$, thus automatically satisfying $\nabla \cdot \mathbf{B} =0$, and the electric current density $\mathbf{j} = (j_x, j_y, 0)=\nabla\times\mathbf{B}/\mu_0$, with its definition 
in terms of velocity moments of the distribution function $f_s$ given by the $x$ and $y$ components of Amp\`{e}re's law,(\ref{eq:ampere-x}) and (\ref{eq:ampere-y}), respectively.

The Vlasov equation, (\ref{eq:vlasov}), is a first order partial differential equation and the characteristics are the equations of motion of a charged particle in the magnetic field $\mathbf{B}$. Due to the symmetries of the problem, one can immediately find three 
constants of motion: namely the Hamiltonian (energy) of a particle of species $s$, $H_s = m_s v^2/2$, and the two
canonical momenta, $p_{xs} = m_s v_{x} + q_s A_x$ and $p_{ys} = m_s v_{y} + q_s A_y$. Any non-negative and differentiable function
$f_s(H_s, p_{xs}, p_{ys})$ with the property that its velocity space moments of all order exist, is then an acceptable solution of 
 (\ref{eq:vlasov}), see \cite{Schindlerbook}. 
 
 The `forward problem' of finding Vlasov-Maxwell equilibria of the type just described is defined by choosing a distribution function $f_s(H_s, p_{xs}, p_{ys})$ for each species $s$ and using it to determine the current density components
 $j_x(A_x,A_y)$ and $j_y(A_x,A_y)$. The final step in completing the solution is solving the two coupled ordinary differential equations (\ref{eq:ampere-x}) and  (\ref{eq:ampere-y}). 
 
If one would like to determine distribution functions for a given magnetic field $\mathbf{B}(z)$, 
one has to solve
the `inverse problem' 
of collisionless plasma equilibrium theory. 
Channell \cite{Channell-1976} has shown that it is possible to formulate this inverse problem as an integral equation using one component of the pressure tensor, which in the case of our choice of coordinates is the $P_{zz}$-component, defined by
\[
P_{zz}(A_x,A_y) = \sum\limits_s m_s\int v_{z}^2\, f_s(H_s,p_{xs}, p_{ys})\, d^3v.
\]
$P_{zz}$ is formally found by taking the moment of $f_s$ by $w_{zs}^2=(v_{z}-u_{zs})^2$, for $u_{zs}$ the fluid velocity in the $z$ direction. However, as in \cite{Channell-1976} we will assume that the form of $f_s$ as a function of its three variables is the same for both particle species and that in particular
\[
f_s(H_s, p_{xs}, p_{ys}) = \frac{n_{0s}}{(\sqrt{2\pi }v_{th,s})^3} \exp\left(-\beta_s H_s\right) g_s (p_{xs}, p_{ys}).
\]
This is an even function of $v_{z}$ and hence $u_{zs}=0$, since
\[
\mathbf{u}_s=\frac{1}{n_s(\mathbf{r},t)}\int \mathbf{v}\,f_s\,d^3v,
\]
for $n_s$ the number density of species $s$. The parameter $\beta_s = (m_sv_{th,s}^2)^{-1}$ normalises the Hamiltonian, with $v_{th,s}$ the \emph{thermal velocity} of species $s$. $n_{0s}$ is a species dependent constant and $g_s$ is an unknown function that has to be determined by solving the inverse problem. This is the main calculation of this paper. $g_s$ will differ between species only due to its dependence on species dependent parameters such as $m_s$, $v_{th,s}$ or $q_s$, but not via its dependence upon the
two canonical momenta $p_{xs}$ and $p_{ys}$. Assuming that the function $P_{zz}(A_x, A_y)$ is known, the inverse problem is defined by the integral equation
\begin{eqnarray}
&&P_{zz}(A_x,A_y)=\frac{\beta_{e}+\beta_{i}}{\beta_{e}\beta_{i}}\frac{n_{0s}}{2\pi {m_s}^2{v_{th,s}}^2}\times\nonumber\\
&&\int_{-\infty}^\infty\int_{-\infty}^\infty  e^{-\beta_{s}\left((p_{xs}-q_sA_x)^2+(p_{ys}-q_sA_y)^2\right)/(2m_s)}g_s(p_{xs},p_{ys})dp_{xs}dp_{ys}, \label{eq:Channell}
\end{eqnarray}
as explained in \cite{Channell-1976}, with examples of solutions in \cite{Moratz-1966,Channell-1976,Harrison-2009b, Wilson-2011,Abraham-Shrauner-2013}. Charge neutrality requires the RHS of (\ref{eq:Channell}) to be independent of species, see \cite{Channell-1976, Harrison-2009b, Wilson-2011, Abraham-Shrauner-2013} for examples and discussion.

In astrophysics, there is a particular interest in \emph{force-free} magnetic fields, see for example \cite{Marshbook}. These are fields such that the electric current density $\mathbf{j}$ is everywhere parallel to $\mathbf{B}$. This can be described mathematically by $\mathbf{j}=\alpha(\mathbf{r})\mathbf{B}$, causing the \emph{Lorentz force} to vanish, $\mathbf{j}\times\mathbf{B}=\mathbf{0}$. The force-free field is said to be nonlinear if $\alpha$ 
varies with position,
linear otherwise and potential if $\alpha=0$. This paper is concerned with a particular nonlinear force-free configuration; the `Force-Free Harris Sheet,' hereafter referred to as FFHS, see \cite{Harrison-2009a,Harrison-2009b,Neukirch-2009,Wilson-2011, Stark-2012},
\[
\mathbf{B}=B_0\left({\rm tanh}(z/L),{\rm sech}(z/L),0\right).
\]
The FFHS describes a \emph{current sheet} - a structure of particular interest in reconnection studies \cite{Priest-2000} - with $L$ the `width' of the current sheet and $B_0$ the constant magnitude of the magnetic field. A force-free equilibrium in Cartesian geometry implies that $P_{zz}$ must have zero spatial gradient (force balance) \cite{Mynick-1979a}. It is shown in \cite{Harrison-2009a} that the $P_{zz}$ must also satisfy Amp\`{e}re's law in the form
\[
\frac{d^2\mathbf{A}}{dz^2}=-\mu_0\frac{\partial P_{zz}}{\partial \mathbf{A}}\label{eq:Amp}.
\]
As shown in \cite{Harrison-2009a}, Amp\`{e}re's law only allows a force-free solution if the function $P_{zz}(A_x, A_y)$
is constant along this solution, i.e. if the solution is also a isocontour of $P_{zz}$ over the $A_x$-$A_y$-plane. In the nonlinear force-free cases found by \cite{Harrison-2009b} and \cite{Abraham-Shrauner-2013} the function $P_{zz}(A_x,A_y)$ had
the form $P_{zz} = P_1(A_x) + P_2(A_y)$ (called the `summative form' from now on).

However, as shown in \cite{Harrison-2009a} and discussed in \cite{Harrison-2009b}, Amp\`{e}re's law admits an infinite number of pressure functions for the same force-free equilibrium. Once a $P_{zz}$ with the correct properties has been found one can define another $P_{zz}$ giving rise to the same current density (for the force-free solution) by using the nonlinear transformation
\begin{equation}
\bar{P}_{zz}=\psi^\prime(P_{ff})^{-1}\psi(P_{zz})\label{eq:Ptrans}.
\end{equation} 
Here any differentiable, non-constant function $\psi$ can be used, such that the RHS is positive, with $P_{ff}$ the pressure, $P_{zz}$, evaluated at the force-free vector potential $\mathbf{A}_{ff}$. 

Obviously, even if the integral equation (\ref{eq:Channell}) can be solved for the original function $P_{zz}(A_x, A_y)$ it is by no means
clear that this is possible for the transformed function $\bar{P}_{zz}$. Usually one would expect that solving (\ref{eq:Channell}) for $g_s$ is much more difficult after the transformation to $\bar{P}_{zz}$. Hence, it is so far unclear whether the transformation property is just a mathematical curiosity or of any practical use.

In this paper we show for the first time that the transformation property of $P_{zz}$ can actually be useful.
As a particularly interesting choice of transformation, this paper deals with exponential transformation
\begin{equation}
\psi(P_{zz})=\exp\left[\frac{1}{P_0}\left(P_{zz}-P_{ff}\right)\right],\label{eq:Pfunc}
\end{equation}
for $P_0$ a positive constant. It is readily seen that $\bar{P}_{zz}|_{\mathbf{A}_{ff}}=P_0$ and so free manipulation of the constant pressure is possible. This is of particular interest because it potentially allows us to freely choose the value of the ratio between
the pressure and the magnetic pressure $B^2/(2 \mu_0)$, the so-called plasma beta. This is not possible in the cases
in which $P_{zz}$ is of summative form, which inevitably have a lower bound of one for the plasma beta.
Note that if $P_{zz}$ is of a `summative form' prior to transformation, then it will be of a `multiplicative form' afterwards, since
\[
\exp(P_1(A_x)+P_2(A_y))=\tilde{P}_1(A_x)\tilde{P}_2(A_y).
\]
The main aim of this paper is to show that there is a mathematically meaningful method of solving (\ref{eq:Channell}) for the
exponentiated form of the pressure function used in \cite{Harrison-2009b} for the force-free Harris sheet.

This paper is organised as follows. In Section 2 we find that by transforming the summative pressure function of \cite{Harrison-2009b} according to (\ref{eq:Ptrans}) and (\ref{eq:Pfunc}), we are able to calculate a new DF, i.e. we solve the inverse problem defined in (\ref{eq:Channell}) - with exponentiated functions on the LHS - for the unknown function $g_s$. We use infinite series of Hermite polynomials in the canonical momenta to do this, and prove that the series are convergent and bounded in Section 3. The necessary matching between the microscopic parameters of the distribution function and the macroscopic parameters of the force-free equilibrium is given in Appendix A. In Appendix B we comment on the non-negativity of the distribution function.

\section{Formal solution to the inverse problem}
The `summative' pressure used in \cite{Harrison-2009b} for a FFHS equilibrium is of the form
\begin{equation}
P_{zz}(A_x,A_y)=\frac{B_0^2}{2\mu_0}\left[\frac{1}{2}\cos\left(\frac{2A_x}{B_0L}\right)+\exp\left(\frac{2A_y}{B_0L}\right)\right]+P_b.\label{eq:Pharrison}
\end{equation}
$P_b>B_0^2/(4\mu_0)$ is a constant that ensures positivity of $P_{zz}$ and $\mu_0$ is the magnetic permeability \emph{in vacuo}. This is the function that we exponentiate according to (\ref{eq:Ptrans}) and (\ref{eq:Pfunc}). Clearly the resultant function will be `multiplicative.' We choose a pressure function and $g_s$ function of the form 
\begin{eqnarray*}
\bar{P}_{zz}&=&n_{0s}\exp\left(-\frac{1}{2\beta_{pl}}\right)\frac{\beta_e+\beta_i}{\beta_e\beta_i}\bar{P}_1(A_x)\bar{P}_2(A_y),\\
g_s&=&\exp\left(-\frac{1}{2\beta_{pl}}\right)g_{1s}(p_{xs})g_{2s}(p_{ys}),
\end{eqnarray*}
such that $\beta_{pl}=2\mu_0P_0/B_0^2$ is the \emph{plasma beta}, a dimensionless constant, and not to be confused with the \emph{thermodynamic beta} $\beta_s$. Separation of variables in (\ref{eq:Channell}) leads to
\begin{eqnarray}
\bar{P}_1(A_x)&=&\frac{1}{\sqrt{2\pi}m_sv_{th,s}}\int_{-\infty}^{\infty}e^{-\beta_{s}\left(p_{xs}-q_sA_x\right)^2/(2m_s)}g_1(p_{xs})dp_{xs},\label{eq:P1tog1}\\
\bar{P}_2(A_y)&=&\frac{1}{\sqrt{2\pi}m_sv_{th,s}}\int_{-\infty}^{\infty}e^{-\beta_{s}\left(p_{ys}-q_sA_y\right)^2/(2m_s)}g_2(p_{ys})dp_{ys}.\label{eq:P2tog2}
\end{eqnarray}  
The pressure tensor components are now written as $1D$ integral transforms of the unknown parts of the distribution function. The Weierstrass transform, $\Phi(x)$ of $\phi(y)$, is defined by
\begin{equation}
\Phi(x)=\mathcal{W}[\phi](x)=\frac{1}{\sqrt{4\pi}}\int_{-\infty}^\infty e^{-(x-y)^2/4}\phi(y)dy,\label{eq:Weier}
\end{equation}
see \cite{Bilodeau-1962}. This is also known as the Gauss transform, Gauss-Weiertrass transform and the Hille transform, \cite{Widder-1951}. Intimately related to the heat/diffusion equation, $\Phi(x)$ represents the temperature profile of an infinite rod one second after it was $\phi(x)$, \cite{Widder-1951}, implying that the Weierstrass transform of a positive function is itself a positive function.
$\bar{P}_1$ and $\bar{P}_2$ are expressed as Weierstrass transforms of $g_{1s}$ and $g_{2s}$ in (\ref{eq:P1tog1}) and (\ref{eq:P2tog2}), respectively, give or take some constant factors. Formally, the operator for the inverse transform is $e^{-D^2}$ with D the differential operator and the exponential suitably interpreted. See \cite{Eddington-1913, Widder-1954} for two different interpretations of this operator. A third and perhaps more computationally `practical' method employs Hermite polynomials, $H_n(x)$, see \cite{Bilodeau-1962}, defined here by a Rodrigues formula;
\[
H_n(x)=(-1)^ne^{x^2}\frac{d^n}{dx^n}e^{-x^2}.
\]
The Weierstrass transform of the $n^{th}$ Hermite polynomial $H_n(y/2)$ is $x^n$. Hence if one knows the coefficients of the Maclaurin expansion of $\Phi(x)$; 
\[
\Phi(x)=\sum_{j=0}^\infty\eta_jx^j,
\]
then the Weierstrass transform can immediately be inverted to obtain the formal expansion
\begin{equation}
\phi(y)=\sum_{j=0}^\infty\eta_jH_j\left(\frac{y}{2}\right).\label{eq:inverseweier}
\end{equation}
For this method to be useful in our problem, the pressure function defined by (\ref{eq:Ptrans}) and (\ref{eq:Pfunc}) must have a convergent Maclaurin expansion, with its coefficients allowing the Hermite series to converge. For the use of Hermite polynomials in Vlasov equilibrium studies see \cite{Grad-1949a,Grad-1949b,Channell-1976,Hewett-1976,Camporeale-2006, Suzuki-2008}. The question of convergence of a formal solution made up of an infinite series of Hermite polynomials was raised in \cite{Hewett-1976}, but not answered. In \cite{Suzuki-2008}, the DF is assumed to be a Maxwellian, multiplied by a sum of Hermite polynomials in velocity and some arbitrary dependence on the vector potential, to be determined later on. Our approach sheds some light on the relationships they derived, whilst beginning from the constants of motion approach. Finally, we note that one of the existing nonlinear force-free Vlasov-Maxwell equilibria known, \cite{Harrison-2009b}, is based on an eigenfunction of the Weierstrass transform, \cite{Wolf-1977}. 

We shall follow a method of the same ilk as Channell, that is to expand the $g_s$ function in Hermite polynomials, with as yet undetermined coefficients. This method is chosen as we believe that it provides the neatest route to deriving `consistency relations'. These consistency relations - derived in Appendix A - are necessary to link the macroscopic parameters that describe properties of the current sheet ($B_0$, $P_0$, $L$) to the microscopic ones that relate to the individual particles ($m_s$, $v_{th,s}$, $q_s$, $n_{0s}$).  
\subsection{The LHS of (\ref{eq:Channell}): Maclaurin expansion of the transformed pressure $\bar{P}_{zz}$}
There is a result from combinatorics due to Eric Temple Bell, that allows one to to extract the coefficients of a power series, $f(x)$, that is itself the exponential of a known power series, $h(x)$, see Bell's original paper \cite{Bell-1934}. If $f(x)$ and $h(x)$ are defined 
\begin{equation}
f(x)=e^{h(x)}, \hspace{5mm} h(x)=\sum_{m=1}^{\infty}\frac{1}{m!}\zeta_mx^m ,\label{eq:exppower}
\end{equation}
then we can use `Complete Bell polynomials', also known as `Exponential Bell Polynomials' and hereafter referred to as CBP's, to write $f(x)$ as
\begin{equation}
f(x)=\sum_{m=0}^\infty \frac{1}{m!}Y_m(\zeta_1,\zeta_2,...,\zeta_m)x^m. \label{eq:Bell}
\end{equation}
$Y_m(\zeta_1,\zeta_2,...\zeta_m)$ is the $m^{\rm{th}}$ CBP. Instructive references on CBP's can be found in \cite{Connon-2010, Kolbig-1994, Comtet-1974, Riordan-1958}. For this paper, the Maclaurin coefficients for the exponential and cosine functions of equation (\ref{eq:Pharrison}) are used as the arguments of the CBPs. These CBPs are used to form the Maclaurin coefficients of $\bar{P}_1$ and $\bar{P}_2$ as in equation (\ref{eq:Bell}). Let us now calculate the Maclaurin expansion of $\bar{P}_1$ using CBPs. First expand the cosine:
\[
\bar{P}_1=\exp\left(\frac{1}{2\beta_{pl}}\cos\left(2\tilde{A}_{x}\right)\right)=\exp\left(\frac{1}{2\beta_{pl}}\right)\exp\left(\sum_{m=1}^\infty\frac{\alpha_m}{m!}\tilde{A}_{x}^m\right)
\]
such that $\alpha_{2l-1}=0$ and $\alpha_{2l}=(-1)^l2^{2l-1}/\beta_{pl}$ for $l\in\mathbf{N}$. $A_x$ is normalised by $B_0L$ to give $\tilde{A}_x$. Now that we have an expression of the form of equation (\ref{eq:exppower}), we can use the CBPs to express $\bar{P}_1$ as a power series in $\tilde{A}_{x}$:
\[
\bar{P}_1=\exp\left(\frac{1}{2\beta_{pl}}\right)\sum_{m=0}^\infty \frac{1}{(2m)!}Y_{2m}(\alpha_1,\alpha_2,...,\alpha_{2m}){\tilde{A}_{x}}^{2m}=\sum_{m=0}^\infty a_{2m}{\tilde{A}_{x}}^{2m},
\]
such that 
\begin{equation}
a_{2m}=\exp\left(\frac{1}{2\beta_{pl}}\right)\frac{1}{(2m)!}Y_{2m}\left(0, -\frac{2}{\beta_{pl}}, 0, \frac{8}{\beta_{pl}}, 0,  ...  ,0 ,\frac{(-1)^m2^{2m-1}}{\beta_{pl}}\right). \label{eq:a2m}
\end{equation}
There is a standard result, \cite{Connon-2010}, of CBPs that $Y_m(a\zeta_1, a^2\zeta_2, ... , a^m\zeta_m)=a^mY_m(\zeta_1, ... , \zeta_m).$ Hence we can simplify equation (\ref{eq:a2m}) to give
\begin{equation}
a_{2m}=\exp\left(\frac{1}{2\beta_{pl}}\right)\frac{(-1)^m2^{2m}}{(2m)!}Y_{2m}\left(0,\frac{1}{2\beta_{pl}}, 0 , ... , 0 , \frac{1}{2\beta_{pl}}\right).\label{eq:a2msimple}
\end{equation} 
The odd terms are all zero and hence not included. This makes sense since $\exp(\cos(x))$ is an even function. 

Next, we find the Maclaurin expansion of $\bar{P}_2(A_y)$ using the same method as above.
\[
\bar{P}_2=\exp\left(\frac{1}{\beta_{pl}}e^{\left(2\tilde{A}_{y}\right)}\right)=\exp\left(\frac{1}{\beta_{pl}}\right)\exp\left(\sum_{n=1}^\infty\frac{\gamma_n}{n!}{\tilde{A}_{y}}^n\right)
\]
such that $\gamma_n=2^n/\beta_{pl}$. $A_y$ is normalised by $B_0L$ to give $\tilde{A}_y$. Making use of equation (\ref{eq:Bell}), we can now express $\bar{P}_2$ as a power series in $\tilde{A}_{y}$:
\begin{equation}
\bar{P}_2=\exp\left(\frac{1}{\beta_{pl}}\right)\sum_{n=0}^\infty \frac{1}{n!}Y_n(\gamma_1,\gamma_2,...,\gamma_n){\tilde{A}_{ys}}^n=\sum_{n=0}^\infty b_n{\tilde{A}_{y}}^n\label{eq:p2exp},
\end{equation}
such that
\[
b_n=\exp\left(\frac{1}{\beta_{pl}}\right)\frac{1}{n!}Y_n\left(\frac{2}{\beta_{pl}}, ..., \frac{2^n}{\beta_{pl}}\right). 
\]
Noticing a geometric factor, we can simplify the CBPs as before to give
\begin{equation}
b_n=\exp\left(\frac{1}{\beta_{pl}}\right)\frac{2^n}{n!}Y_n\left(\frac{1}{\beta_{pl}}, ..., \frac{1}{\beta_{pl}}\right).\label{eq:bnsimple}
\end{equation}

To conclude, the new pressure function $\bar{P}_{zz}$ - calculated by exponentiating $P_{zz}$ as in (\ref{eq:Pharrison}), according to (\ref{eq:Ptrans}) and (\ref{eq:Pfunc}) - has the Maclaurin expansion
\begin{equation}
\bar{P}_{zz}=n_{0s}\exp\left(\frac{-1}{2\beta_{pl}}\right)\frac{\beta_e+\beta_i}{\beta_e\beta_i}\sum_{m=0}^\infty a_{2m}\tilde{A}_x^{2m}\sum_{n=0}^\infty b_n \tilde{A}_y^n, \label{eq:Pmacl}
\end{equation} 
with $a_{2m}$ and $b_n$ defined in (\ref{eq:a2msimple}) and (\ref{eq:bnsimple}) respectively.

\subsection{The RHS of (\ref{eq:Channell}): Hermite expansion of $g_s$}\label{sec:ansatz}
We make the ansatz that $g_s$ is described by the following expansions, in line with the preceding discussion on inversion of the Weierstrass transform;
\begin{eqnarray*}
g_{1s}(p_{xs})&=&\displaystyle\sum_{m=0}^\infty C_{2m,s}H_m\left(\frac{p_{xs}}{\sqrt{2}m_sv_{th,s}}\right), \\
g_{2s}(p_{ys})&=&\displaystyle\sum_{n=0}^\infty D_{ns}H_n\left(\frac{p_{ys}}{\sqrt{2}m_sv_{th,s}}\right),
\end{eqnarray*}
for currently unknown species dependent coefficients $C_{2m,s}$ and $D_{ns}$. We cannot simply `read-off' the coefficients of expansion as in (\ref{eq:inverseweier}), since our integral equations are not quite in the `perfect form' of (\ref{eq:Weier}). Upon computing the integrals of equations (\ref{eq:P1tog1}) and (\ref{eq:P2tog2}) with the above expansions for $g_s$, we have
\begin{eqnarray*}
\bar{P}_1(A_x)&=&\displaystyle\sum_{m=0}^\infty \left(\frac{\sqrt{2}q_s}{m_sv_{th,s}}\right)^{2m}C_{2m,s}\,A_x^{2m},\\
\bar{P}_2(A_y)&=&\displaystyle\sum_{n=0}^\infty \left(\frac{\sqrt{2}q_s}{m_sv_{th,s}}\right)^{n}D_{ns}\,A_y^n,
\end{eqnarray*}
and hence
\begin{eqnarray*}
&&\bar{P}_{zz}(A_x,A_y)=\frac{n_{0s}(\beta_{e}+\beta_{i})}{\beta_{e}\beta_{i}}\exp\left(\frac{-1}{2\beta_{pl}}\right)\\
&&\times\sum_{m=0}^\infty C_{2m,s}\left(\frac{\sqrt{2}q_s}{m_sv_{th,s}}\right)^{2m}A_{x}^{2m}\sum_{n=0}^\infty D_{ns}\left(\frac{\sqrt{2}q_s}{m_sv_{th,s}}\right)^nA_{y}^n.
\end{eqnarray*}
This result appears species-dependent. However, in accordance with \cite{Channell-1976, Harrison-2009b, Wilson-2011} we have to fix the RHS to be both species-independent, and to match with the pressure function that maintains equilibrium with the FFHS, i.e. (\ref{eq:Pmacl}). The results of this calculation are in the Appendix.

We are now able to formally write the DF:
\begin{eqnarray}
&&f_s(H_s,p_{xs},p_{ys})=\frac{n_0\exp\left(\frac{-1}{2\beta_{pl}}\right)}{\left(\sqrt{2\pi}v_{th,s}\right)^3}\exp\left(-\beta_{s}H_s\right)\nonumber\\
&&\times\Bigg[\sum_{m=0}^\infty C_{2m,s}H_{2m}\left(\frac{p_{xs}}{\sqrt{2}m_sv_{th,s}}\right)\sum_{n=0}^\infty D_{ns}H_n\left(\frac{p_{ys}}{\sqrt{2}m_sv_{th,s}}\right)\Bigg]\label{eq:result}
\end{eqnarray}
with $C_{2m,s}$ and $D_{ns}$ given in the Appendix.

\section{Convergence and boundedness of the distribution}
Here we prove that the Hermite series in (\ref{eq:result}) converge. A \emph{sufficient}, general convergence criterion on the coefficients of expansion for $P_{zz}$ is also established, that could be applicable for other series/equilibria. 
\begin{theorem}\label{thm:HermiteConvergence}
The infinite sums 
\[
g_{1s}=\sum_{m=0}^\infty C_{2m,s}H_{2m}\left(\frac{p_{xs}}{\sqrt{2}m_sv_{th,s}}\right)\hspace{3mm}{\rm and}\hspace{3mm}g_{2s}=\sum_{n=0}^\infty D_{ns}H_{n}\left(\frac{p_{ys}}{\sqrt{2}m_sv_{th,s}}\right)
\]
converge $\forall$ $p_{xs}$, $p_{ys}\,\in\mathbb{R}$ and $m_s$, $v_{th,s}\,\in\mathbb{R}^+$, with $C_{2m,s}$ and $D_{ns}$ defined in the Appendix.
\end{theorem}
We shall first prove the convergence of $g_{2s}$ using a combination of the comparison and the ratio test, and then prove convergence of $g_{1s}$ by comparison with $g_{2s}$:
\begin{proof}
Rather than using the CBP formulation of the Maclaurin expansion of the pressure, we shall calculate the expansion `explicitly'. By twice using the Maclaurin expansion of the exponential, we see that
\[
\exp\left(\frac{1}{\beta_{pl}}e^{2\tilde{A}_{y}}\right)=\sum_{n=0}^{\infty}b_n\tilde{A}_y^n, \hspace{3mm} \mbox{\rm s.t.\ } \hspace{3mm}b_n=\frac{1}{n!}\sum_{j=0}^{\infty}\frac{2^jj^n}{j!{\beta_{pl}}^j}.
\]
Using a result from the Appendix we have
\begin{eqnarray}
g_{2s}(p_{ys})&=&\sum_{n=0}^{\infty}D_{ns}H_{n}\left(\frac{p_{ys}}{\sqrt{2}m_sv_{th,s}}\right),\nonumber\\
&=&\sum_{n=0}^{\infty}b_{n}\left(\frac{\delta_s}{\sqrt{2}}\right)^nH_{n}\left(\frac{p_{ys}}{\sqrt{2}m_sv_{th,s}}\right)\label{eq:g2}.
\end{eqnarray}
An upper bound on Hermite polynomials is provided by the identity (see e.g.  \cite{Sansonebook})
\begin{equation}
| H_{j}(x)|<k\sqrt{j!}2^{j/2}\exp\left(x^2/2\right)\hspace{3mm} \mbox{\rm s.t.\ } \hspace{3mm} k=1.086435\, .\label{eq:hermbound}
\end{equation}
This upper bound implies
\[
D_{ns}H_n\left(\frac{p_{ys}}{\sqrt{2}m_sv_{th,s}}\right)<kb_n\delta_s ^n\sqrt{n!}\exp\left(\frac{p_{ys}^2}{4m_s^2v_{th,s}^2}\right).
\]
Working on the level of the series composed of upper bounds, the ratio test requires
\begin{equation}
\lim_{n\to\infty}\Bigg|\frac{b_{n+1}}{b_n}\Bigg|\sqrt{n+1}=\lim_{n\to\infty}\frac{b_{n+1}}{b_n}\sqrt{n+1}<1/\delta_s \label{eq:criterion}
\end{equation}
for a given $\delta_s\in\mathbb{R}^+$. Note that at this stage, we are working with a general $b_n$, and hence equation (\ref{eq:criterion}) is a sufficient criterion for the convergence of a Hermite polynomial series of the type in equation (\ref{eq:g2}), corresponding to a pressure expansion of the type of equation (\ref{eq:p2exp}). The next step is to analyse the asymptotic behaviour of $b_{n+1}/b_{n}$. Explicit expansion of the $b_n$ coefficients gives
\begin{eqnarray*}
b_{n+1}/b_n&=&\frac{1}{n+1}\sum_{j=1}^\infty \frac{2^jj^n}{(j-1)!\beta_{pl}^{j}}\biggr/\sum_{j=1}^\infty\frac{2^jj^n}{j!\beta_{pl}^{ j}}\\
&=&\frac{1}{n+1}\left(\frac{\displaystyle \frac{2}{0!\beta_{pl}}+\frac{2^22^n}{1!\beta_{pl}^{ 2}}+\frac{2^33^n}{2!\beta_{pl}^{ 3}}+...}{\displaystyle \frac{2}{1!\beta_{pl}}+\frac{2^22^n}{2!\beta_{pl}^{ 2}}+\frac{2^33^n}{3!\beta_{pl}^{ 3}}+...}\right)\\
&=&\frac{1}{n+1}\left(\frac{\displaystyle \frac{2}{\beta_{pl}}+2\frac{2^22^n}{2!\beta_{pl}^{ 2}}+3\frac{2^33^n}{3!\beta_{pl}^{ 3}}+...}{\displaystyle \displaystyle \frac{2}{1!\beta_{pl}}+\frac{2^22^n}{2!\beta_{pl}^{ 2}}+\frac{2^33^n}{3!\beta_{pl}^{ 3}}+...}\right)
\end{eqnarray*}
The $k$th `partial sum' of this fraction has the form
\[
r_k=\frac{p_1+2p_2+3p_3+...+kp_k}{p_1+p_2+p_3+...}
\]
with $p_i\asymp 1/i!$, where we write $g\asymp h$ to mean $g/h$ and $h/g$ are bounded away from $0$.
Now since the denominator of the $p_{i}$ increase super-exponentially (factorially) we have $ i p_{i}\asymp p_{i}$ and hence
\[
0<\sum_{i=1}^{\infty}ip_{i}<\infty \hspace{3mm}{\rm and }\hspace{3mm} 0<\sum_{i=1}^{\infty}p_{i}<\infty,
\]
giving $r_{k}\to r_{\infty}\in \mathbb{R}^{+}$ and, more specifically, $r_{\infty}\asymp 1$ in $n$. 
Therefore $b_{n+1}/b_{n}= r_{\infty}/(n+1)\asymp 1/n$. That is to say $b_{n+1}/b_{n}$ grows asymptotically like $1/n$. We then get
\[
\lim_{n\to\infty}\frac{b_{n+1}}{b_n}\sqrt{n+1}=\lim_{n\to\infty}1/\sqrt{n}<1/\delta_s\hspace{3mm}\forall\,\delta_s\in\mathbb{R}^{+}.
\]
Therefore a series with each term at least as large as those of $g_{2s}(p_{ys})$ converges with an infinite radius of convergence ($\forall\, \delta_s$) by the ratio test. Hence $g_{2s}(p_{ys})$ also converges $\forall\, \delta_s$ and $p_{xs}$ by the comparison test. We shall now prove convergence of $g_{1s}$, by comparison with $g_{2s}$.
By explicitly using the Maclaurin expansion of the exponential, and then the power series representation for $\cos^nx$ from \cite{Gradshteyn}:
\begin{eqnarray*}
\cos ^{2n}x&=&\frac{1}{2^{2n}}\left[\sum_{k=0}^{n-1}2{2n \choose k}\cos (2(n-k)x)+{2n \choose n} \right],\\
\cos ^{2n-1}x&=&\frac{1}{2^{2n-2}}\sum_{k=0}^{n-1}{2n-1 \choose k}\cos ((2n-2k-1)x),
\end{eqnarray*}
one can calculate 
\[
\exp\left(\frac{1}{2\beta_{pl}}\cos\left(2\tilde{A}_{x}\right)\right)=\sum_{m=0}^{\infty}a_{2m}\tilde{A}_{x}^{2m}.
\]
The zeroth coefficient is given by $a_0=\exp\left(1/(2\beta_{pl})\right)$, and the rest are
\[
a_{2m}=\frac{2(-1)^m}{(2m)!}\sum_{k=0}^\infty\sum_{j\in J_{k}} \frac{1}{j!(4\beta_{pl})^j}{j \choose k}(j-2k)^{2m}, \label{eq:appendix1}
\]
for $J_{k}=\left\{2k+1, 2k+2, ...\right\}$ and $m\neq 0$. By rearranging the order of summation, $a_{2m}$ can be written 
\[
a_{2m}=\frac{2(-1)^m}{(2m)!}\sum_{j=1}^\infty \frac{1}{j!(4\beta_{pl})^j}\sum_{k=0}^{\lfloor (j-1)/2 \rfloor}{j \choose k}(j-2k)^{2m}, 
\]
where $\lfloor x \rfloor$ is the floor function, denoting the greatest integer less than or equal to $x$. Using a result from the Appendix, we have
\begin{eqnarray*}
g_{1s}(p_{xs})&=&\sum_{m=0}^{\infty}C_{2m,s}H_{2m}\left(\frac{p_{xs}}{\sqrt{2}m_sv_{th,s}}\right),\\
&=&\sum_{m=0}^{\infty}a_{2m}\left(\frac{\delta_s}{\sqrt{2}}\right)^{2m}H_{2m}\left(\frac{p_{xs}}{\sqrt{2}m_sv_{th,s}}\right).
\end{eqnarray*}
Recognising an upper bound in the expression for $a_{2m}$;
\[
\sum_{n=0}^{\lfloor (j-1)/2 \rfloor}{j \choose n}(j-2n)^{2m}\leq j^{2m}\sum_{n=0}^j{j \choose n}=2^{j}j^{2m},
\]
gives 
\begin{eqnarray*}
a_{2m}<\frac{2(-1)^m}{(2m)!}\sum_{j=1}^\infty\frac{2^{j+1}j^{2m}}{j!2^j(2\beta_{pl})^j}&=&2\frac{(-1)^m}{(2m)!}\sum_{j=1}^\infty\frac{j^{2m}}{j!(2\beta_{pl})^j},\\
&\le & \frac{2}{(2m)!}\sum_{j=1}^\infty\frac{j^{2m}}{j!(2\beta_{pl})^j},\\
&=&\frac{1}{(2m)!}\sum_{j=1}^\infty\frac{2^{1-j}j^{2m}}{j!\beta_{pl}^j}<b_{2m}
\end{eqnarray*}
Hence we now have an upper bound on $a_{2m}$ for $m\neq 0$ and we know that $a_{2m+1}=0$, and so is bounded above by $b_{2m+1}$. Note also that $a_0<b_0$. Hence, each term in our series for $g_{1s}(p_{xs})$ is bounded above by a series known to converge $\forall \,\delta_s$ according to
\[
a_l\left(\frac{\delta_s}{\sqrt{2}}\right)^lH_l(x)<b_l\left(\frac{\delta_s}{\sqrt{2}}\right)^lH_l(x).
\]
So by the comparison test, we can now say that $g_{1s}\left(p_{xs}\right)$ is a convergent series. This completes the proof of Theorem~\ref{thm:HermiteConvergence}.
\qedhere
\end{proof}
Note that it is not sufficient for the distribution to be merely convergent. It must also be bounded by a finite constant over all momentum space. Our DF is bounded in momentum space by
\begin{eqnarray*}
f_s&<&e^{-\beta_sH_s}\exp\left(\frac{p_{xs}^2}{4m_s^2v_{th,s}^2}+\frac{p_{ys}^2}{4m_s^2v_{th,ss}^2}\right)S_{1s}S_{2s},\\
&=&e^{-\left(\frac{1}{2}(p_{xs}^2+p_{ys}^2)-2q_s(p_{xs}A_x+p_{ys}A_y)+q_s^2(A_x^2+A_y^2) \right)/(2m_s^2v_{th,s}^2)}S_{1s}S_{2s}
\end{eqnarray*} 
where $S_{1s}$ and $S_{2s}$ are finite constants, since $g_{1s}$ and $g_{2s}$ are convergent. The additional exponential factors come from the upper bounds on Hermite polynomials used above (\ref{eq:hermbound}). This clearly goes to zero for sufficiently large $|p_{xs}|$, $|p_{ys}|$ and is without singularity. We conclude that the distribution is bounded/normalisable.

\section{Summary}
Starting from a one-dimensional, nonlinear force-free \emph{current sheet} equilibrium, we have found a new solution to the inverse problem of calculating non-relativistic, neutral collisionless equilibrium distribution functions for both ions and electrons. This was achieved by transforming a known pressure function by exponentiation, exploiting the fact that an equilibrium pressure function is non-unique. It was noted that by using `Channell's method', the resulting integral equations expressed the pressure as Weierstrass transforms of the unknown parts of the distribution. These were inverted by using a known technique, namely a matched expansion in Hermite polynomials. Complete Bell polynomials were also used to determine the form of the pressure tensor. However, this solution is purely formal, until convergence, boundedness and non-negativity are proven. In Section 3, convergence of $g_s$ was proven by using the ratio and comparison tests, and boundedness was demonstrated. For a brief discussion of the non-negativity of the DF, see Appendix B, and for the full details see \cite{Allanson-2016}.

A significant technical problem has also been overcome, since previous force-free solutions \cite{Harrison-2009b, Abraham-Shrauner-2013, Wilson-2011, Stark-2012} could not describe low-beta configurations, which are prevalent in space and astrophysical plasmas. Whereas the distribution in this paper can describe such plasmas. This could be significant for modelling processes such as reconnection/tearing mode in low-beta environments like the Solar Corona or the Scrape-Off Layer in Tokamaks, see \cite{Biskamp-2000, Fitzpatrick-2007} respectively. Further work in stability analysis and general properties of the DF is a priority.

The spirit of the methods used in this paper are rather general, and could be applied to other magnetic fields with smooth pressure functions $P(A_x,A_y)$, force-free or otherwise: namely the `matching' of Maclaurin and Hermite expansion coefficients; the convergence criteria and the positivity argument. It seems a legitimate project to formally generalise the procedures used in this paper, perhaps extending to include a non-zero electric potential. This could result in a reliable analytic algorithm for calculating valid solutions to the inverse problem for collisionless plasma equilibria, for arbitrary 1D, 2-component magnetic fields. A similarity between solutions of the inverse problem discussed herein, and Green's function solutions of the Diffusion Equation has also been established.

\section*{Acknowledgements}
The authors gratefully acknowledge the financial support of the Leverhulme Trust [F/00268/BB] (TN \& FW), a Science and Technology Facilities Council Consolidated Grant [ST/K000950/1] (TN \& FW), a Science and Technology Facilities Council Doctoral Training Grant [ST/K502327/1] (OA) and an Engineering and Physical Sciences Research Council Doctoral Training Grant [EP/K503162/1] (ST). The research leading to these results has received funding from the European
Commission's Seventh Framework Programme FP7 under the grant agreement SHOCK [284515] (OA, TN \& FW).

\appendix
\section{Consistency of the micro and macroscopic descriptions}
The result of the Maclaurin expansion is to be able to write the pressure tensor in terms of macroscopic parameters as
\[
\bar{P}_{zz}=P_0\exp\left(\frac{-1}{2\beta_{pl}}\right)\sum_{m=0}^\infty a_{2m}\tilde{A}_x^{2m}\sum_{n=0}^\infty b_{n}\tilde{A}_y^{n}
\]
To ensure neutrality $n_i(A_x,A_y)=n_e(A_x,A_y)$,  we require the RHS of equation (\ref{eq:Channell}) to be identical for ions and electrons, see \cite{Channell-1976, Harrison-2009b}. We can let $n_{0s}$ be independent of species, i.e. $n_{0i}=n_{0e}=n_0$. To fix the other parameters of the DF, we insist that the Maclaurin expansion of the pressure tensor above matches up with those found in Section \ref{sec:ansatz}.
\begin{eqnarray*}
\left(\frac{\sqrt{2}q_s}{m_sv_{th,s}}\right)^{2m}C_{2m,s} & = & \left(\frac{1}{B_0L}\right)^{2m}a_{2m}\rightarrow C_{2m,s}=\left(\frac{\delta_s}{\sqrt{2}}\right)^{2m}a_{2m},\\
\left(\frac{\sqrt{2}q_s}{m_sv_{th,s}}\right)^{n}D_{ns} & =  &\left(\frac{1}{B_0L}\right)^nb_n\rightarrow D_{ns}={\rm sgn}(q_s)^n\left(\frac{\delta_s}{\sqrt{2}}\right)^{n}b_{n},
\end{eqnarray*}
and $P_0 =  \frac{n_0(\beta_{e}+\beta_{i})}{\beta_{e}\beta_{i}}$. $\delta_s$ is the species-dependent \emph{magnetisation parameter}, \cite{Fitzpatrickbook}, also used in gyrokinetic theory as the fundamental ordering parameter, \cite{Howes-2006}
\[
\delta_s=\frac{m_sv_{th,s}}{eB_0L}\rightarrow \frac{\delta_e}{\delta_i}=\sqrt{\frac{m_eT_e}{m_iT_i}}.
\]
It is the ratio of the Larmor radius, $\rho_s=v_{th,s}/\Omega_s$, to the characteristic length scale of the system, $L$. $\Omega_s=q_sB_0/m_s$ is the gyrofrequency of particle species $s$ and $T_s$ is the temperature, $k_BT_s=m_sv_{th,s}^2$. When $\delta_s<<1$ then particle species $s$ is highly \emph{magnetised} and a \emph{guiding centre approximation} will be applicable for that species, see \cite{Northrop-1961}. The conditions derived in this Appendix are critical for `fixing' the DF, and making a link between the macroscopic description of the current sheet, with the microscopic one of particles.

\section{Non-negativity of the Distribution}
Since a DF represents a probability density, then it clearly must be non-negative over all phase space. The non-negativity of DFs that are partially represented by sums over Hermite Polynomials has been raised by previous authors, see for example \cite{Abraham-Shrauner-1968,Hewett-1976}. We refer the reader to a recently published article, \cite{Allanson-2016}, for further discussion of non-negativity of $g_s$ functions, represented by (possibly) infinite sums over Hermite Polynomials. To paraphrase the results therein, it is claimed that -- provided $g_s$ is differentiable and convergent -- the $g_s$ function will be positive for a sufficiently magnetised plasma, i.e. for $0<\delta_s<\delta_c\le \infty$, for some critical value of the magnetisation parameter, $\delta_c$.

\bibliographystyle{plain}


\end{document}